\documentclass[twocolumn,pra,showpacs,aps,superscriptaddress]{revtex4}

\usepackage{amsmath}
\usepackage{amssymb}
\usepackage{latexsym}
\usepackage{graphicx}
\usepackage{float}
\usepackage{color}
\begin{document}
\title{Noise Correlation Scalings: Revisiting the Quantum Phase Transition in
Incommensurate Lattices with Hard-Core Bosons}

\author{Kai He}
\affiliation{Department of Physics, Georgetown University, Washington, DC, 20057, USA}
\affiliation{Joint Quantum Institute, National Institute of Standards and Technology,
Gaithersburg, Maryland 20899, USA}
\author{Indubala I. Satija}
\affiliation{Joint Quantum Institute, National Institute of Standards and Technology,
Gaithersburg, Maryland 20899, USA}
\affiliation{Department of Physics, George Mason University, Fairfax, Virginia, 22030, USA}
\author{Charles W. Clark}
\affiliation{Joint Quantum Institute, National Institute of Standards and Technology,
Gaithersburg, Maryland 20899, USA}
\author{Ana Maria Rey}
\affiliation{JILA, National Institute of Standards and Technology, and
University of Colorado, Boulder, Colorado 80309, USA}
\author{Marcos Rigol}
\affiliation{Department of Physics, Georgetown University, Washington, DC, 20057, USA}

\begin{abstract}
Finite size scalings of the momentum distribution and noise correlations are performed to
study Mott insulator, Bose glass, and superfluid quantum phases in hard-core bosons (HCBs)
subjected to quasi-periodic disorder. The exponents of the correlation functions at the
Superfluid to Bose glass (SF-BG) transition are found to be approximately one half of the ones
that characterizes the superfluid phase. The derivatives of the peak intensities
of the correlation functions with respect to quasiperiodic disorder are shown to diverge
at the SF-BG critical point. This behavior does not occur in the corresponding free fermion
system, which also exhibits an Anderson-like transition at the same critical point, and thus
provides a unique experimental tool to locate the phase transition in interacting bosonic
systems. We also report on the absence of primary sublattice peaks in the momentum distribution
of the superfluid phase for special fillings.
\end{abstract}
\pacs{03.75.Hh, 05.30.Rt, 05.30.Jp, 42.50.Lc}
\maketitle

\section{Introduction}

Since the celebrated work of Anderson \cite{Anderson58PR} and others about half a century ago,
the subject of disorder and quantum phase transitions induced by disorder continues to attract
interest \cite{reviewdisorder}. In spite of being intensively studied, the
interplay between interactions and disorder remains an open frontier.  In recent years,
a new impetus toward the understanding of disorder systems has emerged in the context of ultracold
atomic systems because of their extraordinary degree of tunability. In those systems, it is now possible
to control the degree of disorder either by imprinting speckle patterns \cite{Billy08Nature,Lye05PRL}
or by superimposing a secondary lattice on the main lattice to generate a quasiperiodic potential
\cite{Lye07PRA, Fallani07PRL, Guarrera08PRL, Roati08Nature}. Using those techniques,
Anderson localization has already been observed in one dimension \cite{Roati08Nature, Lye07PRA, Billy08Nature}.
At the same time, the possibility of exploring strongly interacting regimes with ultra-cold atoms has been
demonstrated by observation of the superfluid to Mott insulator transition
\cite{Greiner02Nature,stoferle_moritz_04,spielman_phillips_07,spielman_phillips_08}, and the
realization of the Tonks-Girardeau gas, i.e., a gas of impenetrable (hard-core) bosons (HCBs),
in one dimension \cite{Paredes04Nature,Kinoshita04Science,Kinoshita05PRL}. The combined
investigation of the effect of disorder and correlations in low-dimensional systems is therefore a window
that is now open for experimental exploration \cite{review1D} and has driven many recent investigations
\cite{Rey06PRA,Scarola06PRArc,Shuming2010PRE,Cai2010PRA}.

In this paper we revisit the problem of one-dimensional (1D) HCBs with pseudo-random disorder generated
by imposing two-color lattices with periods that are mutually incommensurate. Even in the absence of
interactions, one-dimensional quasiperiodic systems exhibit a localization-delocalization transition.
In contrast to truly random 1D systems where states are localized irrespective of the strength
of disorder, quasiperiodic systems exhibit a quantum phase transition from extended to localized at a
finite strength of the quasiperiodic disorder. In the interacting soft-core regime,
several studies have addressed the phases that result from the interplay between pseudorandom disorder
and strong correlations. Those studies have utilized different computational techniques such as
exact diagonalization \cite{Roth03PRA, Louis07JLTP}, quantum Monte Carlo simulations \cite{Roscilde08PRA},
and the density-matrix renormalization group method \cite{Roux08PRA, Deng08PRA}. A recent
comprehensive discussion of these, and other, results can be found in Ref.~\cite{review1D}. In the present
work, we perform an exact numerical study of ultracold bosons in the hard-core regime, as well as
spinless fermions, and investigate the effects of the interplay between disorder and interactions in the
momentum distribution and noise correlations of those systems. These two observables can be
experimentally probed through time-of-flight measurements \cite{Altman04PRA,Folling05Nature}.

Noise correlations have been examined before for several HCB systems, including homogeneous, period-2 superlattices,
and truly random disordered systems \cite{Rey06JPB, Rey06NJP, He2011PRA}. However, previous
studies of HCBs in incommensurate superlattices \cite{Rey06PRA} were limited to small system sizes,
which prevented an accurate determination of the scaling of off-diagonal correlations at the superfluid
to insulator transition. The present study is partially motivated by an alternative approach
to computing four-point correlations \cite{He2011PRA}. This approach enables the investigation of
larger systems and opens the possibility of a detailed scaling analysis and better characterization of the
quantum phase transition. One of our key results here is the determination of the exponents
at the transition point. Furthermore, as one approaches the thermodynamic limit, roots of divergent
behavior in the derivative of the correlation peak intensities become apparent at the HCB
localization-delocalization phase transition. That singularity is absent in the free fermionic
system. This is another important finding in this work.

The paper is structured as follows. Following the background introduction, Sec.\ \ref{sec:model} describes
the model and defines the correlation functions. The localization transition is then reviewed in
Sec.\ \ref{sec:local}. Section \ref{sec:scaling} is devoted to the scaling study of momentum distribution
and noise correlations, revealing the distinctive behavior at the critical point.
In Sec.\ \ref{sec:transition}, the dependence of both quantities on the strength of disorder is explored
and a new characterization of the phase transition is given. Section \ref{sec:absence} reports on the
peculiar absence of correlation features for some fillings. In Sec.\ \ref{sec:conclusions}  we summarize
our results.

\subsection{Model}\label{sec:model}

One-dimensional hard-core bosons in incommensurate superlattices, within the one-band approximation,
are described by the following Hamiltonian:
\begin{equation}\label{eq:Ham}
    \hat{H}_\textrm{HCB}=-t\sum_j (\hat{b}^\dagger_j \hat{b}_{j+1}+\textrm{H.c.})+ \sum_j V_j\hat{n}^b_j,
\end{equation}
where $\hat{b}^\dagger_j(\hat{b}_j)$ is the bosonic creation (annihilation) operator at the site $j$,
satisfying the bosonic commutation relations $[\hat{b}_i, \hat{b}^\dagger_j]=\delta_{ij}$ and the
on-site constraints $\hat{b}^{\dagger 2}_{i}= \hat{b}^2_{i}=0$; $\hat{n}^b_j=\hat{b}^\dagger_j\hat{b}_j$
is the bosonic particle number operator. Here $t$ is the hopping energy between adjacent sites,
which we set to one $(t=1)$ in our work, and $V_j$ denotes the external potential on site $j$ generated by the
secondary lattice. Note that the primary lattice is the only one that is described within the tight binding
approximation. $V_j$, on the other hand, has the following form:
\begin{equation}\label{eq:QPpotential}
    V_j=2\lambda \cos(2\pi\sigma j+\delta),
\end{equation}
where the parameter $\lambda$ is proportional to the intensity of the lasers used to create the secondary
lattice, $\sigma$ is the ratio between the wave vectors of the two lattices, and $\delta$ is a generic
phase factor. Here, we choose $\delta$ to be zero without lost of generality. Finally, we note that periodic
boundary conditions are considered throughout this work.

Quasiperiodic lattices are characterized by an irrational $\sigma$, which results from the incommensurate
periodicity of the two superimposed lattices. In our studies, we choose $\sigma$ to be the reverse golden ratio,
$\sigma=(\sqrt{5}-1)/2\doteq0.61803...$. This choice is motivated by the fact that the golden mean
is considered to be the most irrational number \cite{Sokoloff85, Shuming2010PRE} and has simple
number theoretical properties. In our numerical calculations, $\sigma$ is approximated by the ratio of
two consecutive Fibonacci numbers $F_{M-1}/F_M$, where the Fibonacci numbers are defined by the
recursion relation $F_{i+1}=F_i+F_{i-1}$, with $F_1=F_0=1$. Within the approximation followed in this manuscript, model \eqref{eq:Ham} on a quasiperiodic superlattice reduces to a model on a periodic superlattice  whose properties are well known ($\sigma$ is rational). The latter nevertheless captures the localization-delocalization transition exhibited by the quasi-periodic system \cite{Buonsante04PRA, Rousseau06PRB}.

We focus our analysis on two observables, the momentum distribution function $n_k$ and the noise
correlations $\Delta_{kk'}$, defined as
\begin{eqnarray}
\label{eq:defmom}   n_k&=&\frac{1}{L} \sum_{i j} e^{i ka(i-j)}
\langle\hat{b}^\dagger_i \hat{b}_j\rangle, \\
\label{eq:defnoi} \Delta_{k k^\prime} &\equiv& \langle \hat{n}_k
\hat{n}_{k^\prime} \rangle \ - \langle \hat{n}_k\rangle \langle
\hat{n}_{k^\prime}\rangle - \langle \hat{n}_k\rangle \
(\delta_{k-k^\prime, nK}-\delta_{k,k^\prime}),\notag
\end{eqnarray}
where $k(k')$ denote the momentum values, $L$ the number of lattice sites, $a$ the lattice constant,
$K=2\pi/a$ the reciprocal lattice vector, and $n$ an integer. The expression for $\Delta_{k k^\prime}$
was derived in Ref.~\cite{Rey06JPB}. The linear term, proportional to  $\langle \hat{n}_k\rangle$,
has its roots in the commutation relations of bosonic fields and the requirement of using normal ordered operators
if one wants to map time-of-flight observables to {\it in-situ} correlations evaluated before the expansion
and restricted to the lowest band. See Ref.~\cite{Toth2008} for details.

An important point to be kept in mind when evaluating it is that, for hard-core bosons, only normal
ordered expectation values can be computed using the mapping to
a spin-1/2 chain and then to noninteracting fermions \cite{Rey06JPB,He2011PRA}. In contrast to earlier
studies where correlations were calculated using Wick's theorem \cite{Rey06JPB}, here we follow an
alternative procedure based on Slater determinants \cite{Marcos04PRA2,Marcos05PRA,He2011PRA}, which,
for completeness, is briefly reviewed in Appendix \ref{app:method}.

\subsection{Localization transition}\label{sec:local}

As mentioned before, two-color superlattices characterized by an irrational $\sigma$,  are intermediate between periodic and fully random, and exhibit a metal-insulator
transition in one dimension. The critical point $\lambda_c=t$ describes the onset of localization.
Below criticality ($\lambda <\lambda_c$), single-particle states are extended, namely, they are
Bloch-like states. Above the critical point, single-particle states are exponentially localized
with localization length $\xi = \ln(\lambda)^{-1}$ \cite{Aubry80}.

\begin{figure*}[!ht]
\centering
  \includegraphics[width=0.70\textwidth]{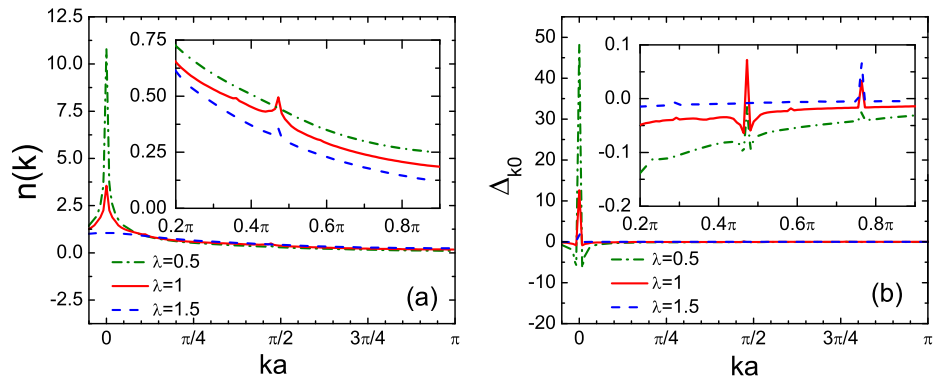}
\vspace{-0.3cm}
  \caption{(Color online) (a) Momentum distribution $n(k)$ and (b) noise correlations $\Delta_{k0}$, of HCBs,
as a function of $k$ for three different values of $\lambda$. In all cases $L=233$, $N=166$, and $\sigma=144/233$.
The insets in (a) and (b) show the set of sublattice peaks in the momentum distribution and noise correlations, respectively.}
\label{fig:profile}
\end{figure*}

In the extended phase, the energy spectrum consists of a set of bands (it becomes a Cantor set
at the localization threshold). The gaps are located at $k=\pm n\sigma$(mod 1)$\pi/a$ in the first
Brillouin zone, indexed by an integer $n$. Note that as $n$ increases the width of the gaps narrows.
As a consequence of our selection of $\sigma=F_{M-1}/F_M$, the single-particle spectrum
consists of $F_M$ bands and $F_M-1$ gaps. The filling factor $\rho=N/L$ in a many-body system,
where $N$ is the number of particles and  $L$ the number of sites, is then a control parameter that
tunes transitions between different phases. For fillings $\rho=n\sigma$(mod 1) and
$\rho=1-n\sigma$(mod 1), the ground state of HCBs in the extended regime is a Mott insulator. This
insulating state is incompressible and it is also referred to as an incommensurate insulator
\cite{Rey06PRA, Shuming2010PRE, Roscilde08PRA}. The fact that the system is insulating at those
fillings can be understood because the conduction bands in the fermionic system, to which the HCBs can be mapped,
are filled \cite{Rey06PRA, Shuming2010PRE}. For all other fillings, the ground state is superfluid.
To be able to observe the fractional Mott insulating phases, the lattice size must be identical to
(or multiple integers of) the period of the incommensurate potential, which means that we choose $L=F_M$
while having $\sigma=F_{M-1}/F_M$. In our study, we consider lattices with $L=$34,55,89,144, and 233.
For $\lambda > \lambda_c$, i.e., in the localized regime, the ground state remains an
incompressible insulator for fillings $\rho=n\sigma$(mod 1) and $\rho=1-n\sigma$(mod 1). For all
other fillings, on the other hand, the ground state is a Bose glass. The Bose-glass phase is a compressible insulating state
where the absence of transport is the result of localization \cite{review1D}.

\section{Quantum correlations: Scaling Analysis}\label{sec:scaling}

In this section, we perform a finite size scaling study of the momentum distribution as well as the noise correlations
in different regimes.  Typical momentum profiles and noise correlation patterns are shown in
Fig.\ \ref{fig:profile}. They were calculated on a system with 200 sites at half filling and for three different
values of $\lambda$ representing the superfluid phase, the critical point, and the localized phase.

The momentum distribution reflects the quantum coherence of the system. It exhibits a sharp diffraction
pattern in the extended regime (superfluid phase) and a flat profile in the Bose glass and Mott insulating phases [
see Fig.~\ref{fig:profile}(a)]. The interference pattern is characterized by a peak at $k=0$ and additional
sublattice peaks at the corresponding reciprocal lattice vectors. In the quasiperiodic lattice under consideration, they
are found at $ka=\pm n\sigma$(mod 1)$2\pi$ and $ka=\pm(1- n\sigma)$(mod 1)$2\pi$. In general, only peaks associated
with the main reciprocal lattice vectors, $n=1$, are distinguishable. The $k=0$ peak is the sharpest peak and signals
the superfluid character of the extended phase.

The noise correlation pattern contains information of the momentum-momentum correlations and is also known as
Hanbury Brown--Twiss interferometry. It exhibits interference peaks in all phases, even in the insulating ones.
In contrast to the momentum distribution, the noise correlation fringes have their root in the quantum statistics
of the particles. For bosonic systems, the peaks are positive and are a manifestation of bunching. For fermionic
systems, they are negative and signal antibunching \cite{Altman04PRA,Folling05Nature}.

The height of the noise correlation peaks and their underlying background significantly vary in the various quantum
phases studied here. For example, a negative background is found in the superfluid regime. Also, the sublattice peaks
disappear in the Mott insulting phase which occur at unit filling. The insets in Figs.~\ref{fig:profile}(a) and
\ref{fig:profile}(b) show the most noticeable sublattice peaks in our systems.

\begin{figure}[!t]
\centering
  \includegraphics[width=0.45\textwidth]{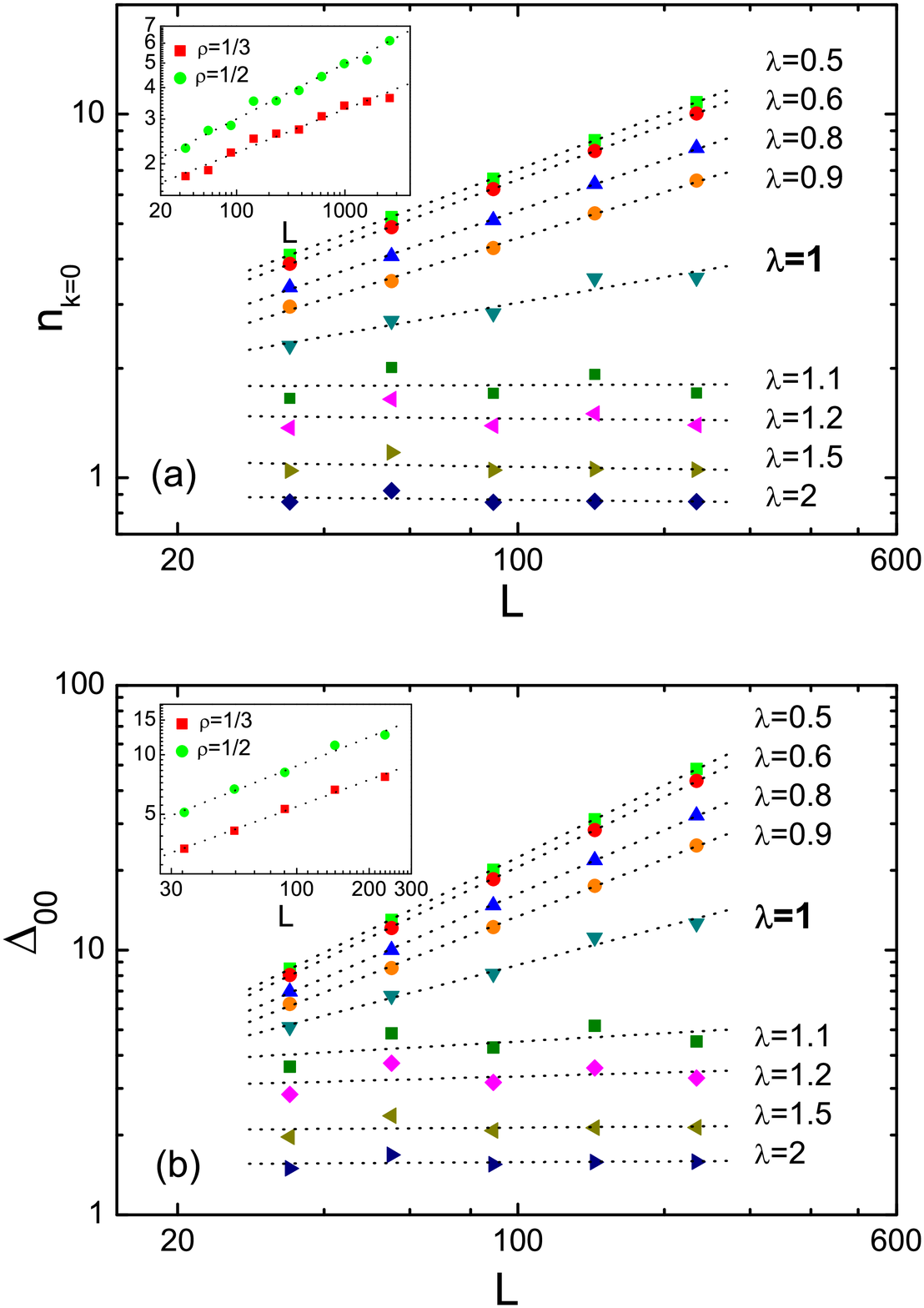}
\vspace{-0.3cm}
   \caption{(Color online) Scaling of (a) $n_{k=0}$ and (b) the central peak $\Delta_{00}$ in half-filled systems for various
    values of $\lambda$. The inset in (a) depicts the scaling of $n_{k=0}$ for $\lambda=\lambda_c$ and up to larger
    system sizes. The black dotted lines are power-law fits to the data, from which the exponents (see text)
    indicate the distinction between the extended regime, the transition point, and the localized regime.\label{fig:sizescal1}}
\end{figure}

In Fig.\ \ref{fig:sizescal1} we plot the central peak of the momentum distribution function $n_{k=0}$ (a) and the
noise correlation peak $\Delta_{00}$ (b) as a function of the system size $L$ for various values of $\lambda$. Note
that the filling factor is fixed at $\rho=0.5$ so that a superfluid state is associated with the $\lambda<1$ cases.
The distinction in the scaling behavior in the extended regime, at the transition point, and in the localized regime
is apparent. One can see that both $n_{k=0}$ and $\Delta_{00}$ exhibit power-law scaling in the superfluid phase
as well as at criticality, while in the Bose-glass phase they remain essentially unchanged with changing $L$.

In Fig.\ \ref{fig:sizescal1}(a), the four sets of values of $n_{k=0}$ vs $L$ for $\lambda=$0.5,0.6,0.8, and 0.9 exhibit
exponents $0.503\pm0.001$, $0.496\pm0.002$, $0.462\pm0.007$, and $0.42\pm0.01$, respectively, which are compatible with
the well-known $n_{k=0} \sim \sqrt{L}$ scaling for HCBs in the superfluid regime, where the quasi-long-range order is
present in the one-particle correlations. In Fig.\ \ref{fig:sizescal1}(b), the exponents for $\Delta_{00}$ in the extended
regime are $0.903\pm0.003$, $0.878\pm0.004$, $0.799\pm0.007$, and $0.72\pm0.01$, also for $\lambda=$0.5, 0.6, 0.8, and 0.9,
respectively. These results are compatible with the linear behavior $\Delta_{00} \sim L$ reported in Ref.~\cite{He2011PRA},
where the superfluid phase of the homogeneous system and the period-2 lattice was studied. Overall, the results in
the extended regime are consistent with the fact that for $\lambda<1$ the system is a Luttinger liquid with $K=1$
\cite{review1D}. On the other hand, for $\lambda>1$, Fig.\ \ref{fig:sizescal1} shows that both $n_{k=0}$ and $\Delta_{00}$
do not scale with system size, which is a demonstration of the presence of exponentially decaying correlations in the
Bose-glass phase, and hence the lack of Luttinger liquid behavior.

The behavior at the critical point is the most interesting one, as the system
is not expected to be described by the Luttinger liquid theory. Still, we find
a power-law scaling of both $n_{k=0}$ and $\Delta_{00}$ with an exponent that
is roughly one-half of the one in the Luttinger liquid regime. The inset in
Fig.\ \ref{fig:sizescal1}(a) shows $n_{k=0}$ for $\lambda=1$ with larger
system sizes than those depicted in the main panel and results for an
additional filling $\rho=1/3$. We find that $n_{k=0} \sim L^{0.22\pm0.01}$ at
$\rho=1/2$ and $n_{k=0} \sim L^{0.17\pm0.01}$ at $\rho=1/3$. In Fig.\ \ref{fig:sizescal1}(b),
$\Delta_{00}$ at the transition point scales with an exponent of $\sim 0.48\pm0.03$ at
$\rho=1/2$ and $\sim 0.45\pm0.02$ at $\rho=1/3$. Notice that there are apparent
fluctuations in the data, at the critical point and above, with increasing system
size. These are finite size effects whose main origin may be the fact that the
filling factor, as well as $\sigma$, fluctuate
slightly between lattices with different size. Unfortunately, the number of particles
and the number of lattice sites cannot be accommodated to give the same exact filling
factor with increasing system size. Still, our results are consistent with a halving
of the exponents from their Luttinger liquid values. This is something that deserves
further theoretical investigation.

\begin{figure}[!t]
  \centering
  \includegraphics[width=0.45\textwidth]{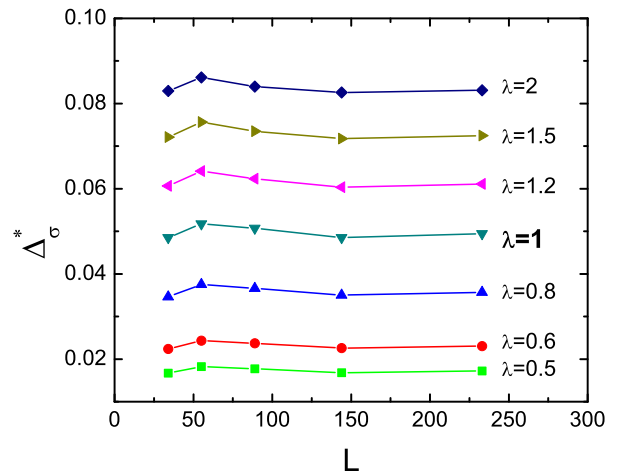}
\vspace{-0.3cm}
  \caption{(Color online) Scaling of the sublattice peaks $\Delta^\ast_\sigma$ of half-filled systems for various $\lambda$. These plots
  show that the sublattice peaks do not scale with system size.}\label{fig:sizescal2}
\end{figure}

We carry out a similar study for the sublattice peaks. The $L$ dependence of the primary
sublattice peak ($n=1$) is presented in Fig.\ \ref{fig:sizescal2} for the same
values of $\lambda$ as in Fig.~\ref{fig:sizescal1}. Notice that the sublattice
peaks are immersed in a negative background, as shown in the inset Fig.\ \ref{fig:profile}(b), so we quantify their height by
subtracting the background; that is, we define $\Delta^\ast_\sigma \equiv
\Delta_{\sigma \frac{2\pi}{a},0}-\frac{1}{2}(\Delta_{\sigma \frac{2\pi}
{a}+\delta k,0}+\Delta_{\sigma \frac{2\pi}{a}-\delta k,0})$, with $\delta
k=2\pi/La$. For systems at $\rho=0.5$, the sublattice peaks do not scale with
system size for any value of $\lambda$. This finding is surprising for
$\lambda<\lambda_c$ because, as follows from the scaling of the $\Delta_{00}$
peak, quasi-long-range order is present in the system in that regime. On the
other hand, the absence of those peaks is expected in the glassy phase where
correlations decay exponentially. Further analysis reveals that, within a
range of filling factors from $\rho=1-\sigma$ to $\rho=\sigma$,
$\Delta^\ast_\sigma$ does not scale with system size in the superfluid phase.
This will be discussed in detail in Sec.\ \ref{sec:absence}.

\begin{figure*}[!ht]
  \centering
  \includegraphics[width=0.67\textwidth]{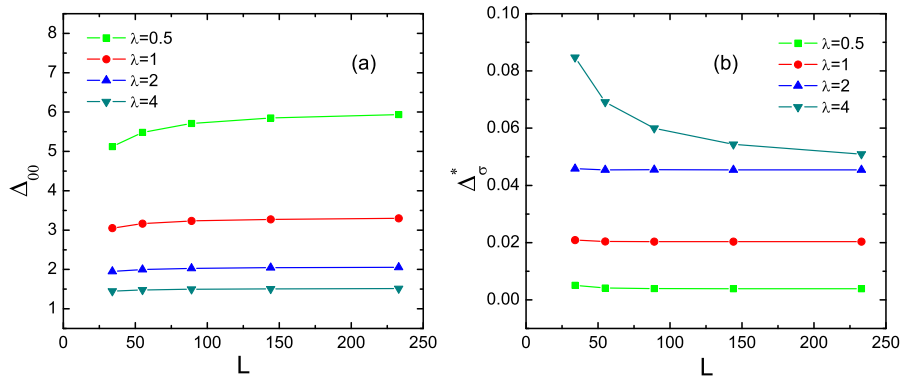}
\vspace{-0.3cm}
  \caption{(Color online) (a) Central noise correlation peak $\Delta_{00}$ and (b) the sublattice peak $\Delta^\ast_\sigma$
  as a function of $L$ in systems at incommensurate filling $\rho=\sigma$.}\label{fig:sizescal3}
\end{figure*}

At the incommensurate fillings $\rho=n\sigma$(mod 1) and $\rho=1-n\sigma$(mod 1), a Mott insulating
phase rather than a superfluid one is present for all $\lambda\neq0$. We then study the noise correlation peaks
in systems with incommensurate filling $\rho=\sigma$. In Fig.\ \ref{fig:sizescal3}, both $\Delta_{00}$
and $\Delta^\ast_\sigma$ are plotted as a function of $L$ for various values of $\lambda$. The absence of
scaling is apparent in both noise correlation peaks for all regimes. For those fillings, $n_k$ and the noise
correlations do not distinguish the localization delocalization transition that occurs at $\lambda_c$.

\section{Quantum Phase Transition}\label{sec:transition}

In order to gain a better understanding of the disorder-induced phase transition and its signatures in our
strongly interacting system, we study the behavior of the noise correlations in the vicinity of the critical point.
Our study involves monitoring the peak intensities and their derivatives as the strength of the quasiperiodic order
is tuned.

\begin{figure*}[!ht]
    \centering
    \includegraphics[width=0.7\textwidth]{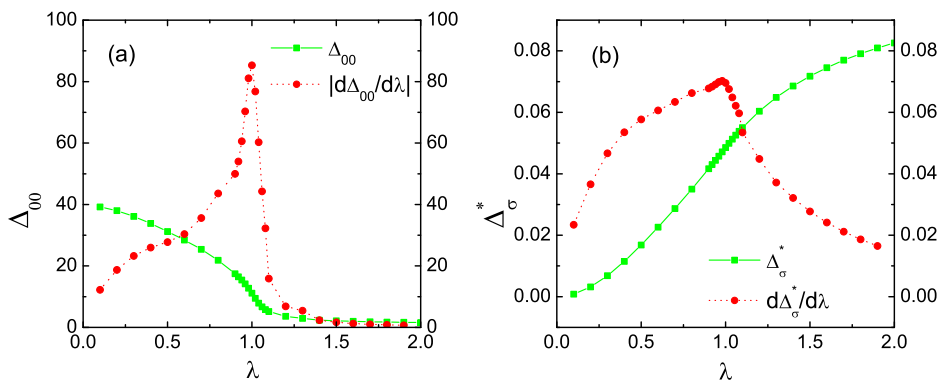}
\vspace{-0.3cm}
    \caption{(Color online) (a) Central peak $\Delta_{00}$ and (b) sublattice peak $\Delta^\ast_\sigma$ as a function of
    $\lambda$ and their corresponding derivatives in a system with $N=72, L=144$. The onset of the superfluid
    to Bose-glass transition is marked with a peak in the first-order derivative of both noise correlations.
    Note that in both panels the left scale applies to the intensity of the peak while the right one applies to its derivative.}
    \label{fig:amphcb}
\end{figure*}

Figures \ref{fig:amphcb}(a) and \ref{fig:amphcb}(b) depict the results for the central peak and the primary
sublattice peak, respectively, as a function of $\lambda$ in a half-filled system. With the increase of disorder,
the height of the central peak ($\Delta_{00}$) decreases and attains an asymptotic value $\Delta_{00}=\rho(\rho+1)$
as $\lambda\rightarrow\infty$ \cite{Rey06PRA,He2011PRA}, while the height of the sublattice peak increases with disorder
and saturates at a finite but small value as $\lambda\rightarrow\infty$. In the neighborhood of the the critical point,
a kink is seen in both peaks as $\lambda$ is varied. In order to better understand the behavior in the critical region,
we compute $d\Delta_{00}/d\lambda$. Figure \ref{fig:amphcb}(a) shows that the derivative of the height of the
central peak exhibits a maximum at the transition point. A similar feature is seen in the derivative of the primary
sublattice peak intensity [Fig.~\ref{fig:amphcb}(b)].

It is instructive to compare the behavior of the noise correlation peaks at the localization transition in HCB systems
with the one at the metal-insulator transition in the corresponding free fermionic system. Analogous to the HCB case,
in an incommensurate lattice system, noninteracting spinless fermions exhibit a localization transition at $\lambda_c$.
Fingerprints of the phase transition in the correlation functions of the fermionic system have been systematically
studied in Ref.\ \cite{Shuming2010PRE}. Our comparative study is aimed at singling out distinctive features in both systems.
Notice that the HCBs are interacting and have bosonic statistics, while their fermionic counterpart is noninteracting
and, of course, has fermionic statistics. Noise correlations are sensitive to both interactions and the statistics
of the particles involved.

\begin{figure*}[!ht]
  \includegraphics[width=0.70\textwidth]{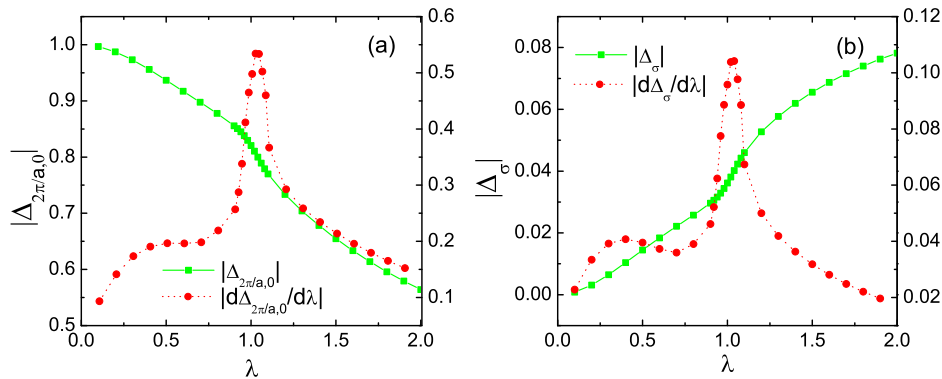}
\vspace{-0.3cm}
  \caption{(Color online) (a) Primary dip $|\Delta_{\frac{2\pi}{a},0}|$ and its derivative, and (b) the sublattice dip $|\Delta_\sigma|$
  and its derivative, are plotted as a function of $\lambda$ for noninteracting fermions. $N=72,\ L=144$. In the fermionic
  system the critical point is also signaled by sharp peaks in the first-order derivatives. Note that again in both panels
  the left scale applies to the intensity of the dip while the right one applies to its derivative.}
  \label{fig:ampk0k1fer}
\end{figure*}

Figure \ref{fig:ampk0k1fer} shows the behavior of noise correlations in the fermionic system vs $\lambda$. In this case,
distinctive features of the noise correlations are the Bragg dip at $ka=2\pi$ [depicted in Fig.~\ref{fig:ampk0k1fer}(a)]
and the primary sublattice dip at $ka=2\sigma\pi$ [depicted in Fig.~\ref{fig:ampk0k1fer}(b)], as well as their derivatives
with respect to $\lambda$. As in the HCB case, the metal-insulator transition in the fermionic system is also signaled
by peaks in the first-order derivatives of noise correlations at the critical point.

\onecolumngrid

\newpage

\twocolumngrid

\begin{figure}[!b]
  \centering
  \includegraphics[width=0.43\textwidth]{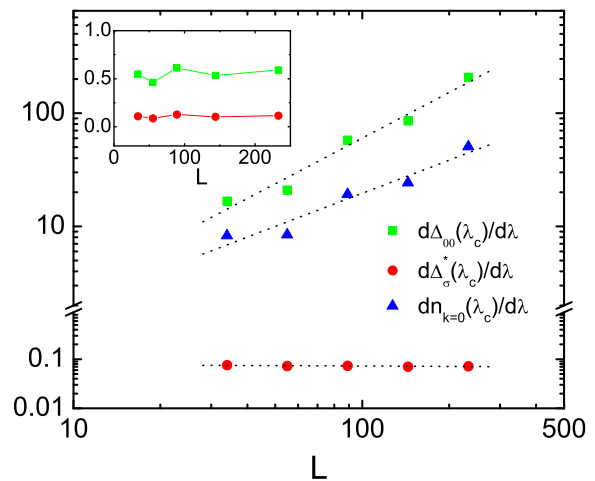}
\vspace{-0.3cm}
  \caption{(Color online) (Main panel) Derivatives of $\Delta_{00}, \Delta^\ast_\sigma$, and $n_{k=0}$ as a function of the system size
  at $\lambda_c$ in HCB systems. A divergence in both $d\Delta_{00}(\lambda_c)/d\lambda$ and $n_{k=0}(\lambda_c)/d\lambda$
  is seen as the system size increases, while $d\Delta^\ast_\sigma(\lambda_c)/d\lambda$ saturates to a constant value.
  The inset shows $|d\Delta_{\frac{2\pi}{a},0}/d\lambda|$ and $|d\Delta_\sigma/d\lambda|$ as a function of $L$, and also at
  $\lambda_c$, in fermionic systems. No size dependence is seen in this case. The mean occupation per lattice site is
  $\rho=0.5$ for both cases.}
  \label{fig:sizescalderv}
\end{figure}

A study of the scaling of the peaks and their derivatives with increasing system size makes evident that there is an
important difference between their behavior for HCBs and fermions. The results of such a scaling, for the derivatives at the
critical point, are presented in Fig.\ \ref{fig:sizescalderv}. For HCB systems, we plot
$d\Delta_{00}(\lambda_c)/d\lambda$ and $d\Delta^\ast_\sigma(\lambda_c)/d\lambda$ as a function of $L$.
One can clearly see that $d\Delta_{00}(\lambda_c)/d\lambda$ diverges with increasing system size. This is
expected as, on the left of the transition point, $\Delta_{00}$ increases with system size while, on the right of
the transition, $\Delta_{00}$ saturates to a $\lambda$-dependent (size-independent) value as the system size is increased.
As also shown in the figure, the behavior of $dn_{k=0}(\lambda_c)/d\lambda$ is qualitatively similar;
$d\Delta^\ast_\sigma(\lambda_c)/d\lambda$,
on the other hand, does not change with system size, as follows from the lack of scaling of this sublattice peak
discussed with Fig.~\ref{fig:sizescal2}. For the fermions,  shown in the inset in Fig.\ \ref{fig:sizescalderv},
one can see that none of the derivatives scales with system size.

Hence, in the thermodynamic limit, the derivatives of both the HCB and fermion noise correlations peak heights exhibit maxima
at the critical point. However, it is only in the HCB system that these maxima diverge. Hence, they provide very sharp
signatures to locate the phase transition in experiments with ultracold gases in the bosonic case.

\section{Absence of primary sublattice peaks}\label{sec:absence}

In this section, we analyze the absence of scaling of the primary sublattice peaks, in the superfluid phase,
for a specific window of filling factors.

\begin{figure}[tb]
    \centering
  \includegraphics[width=0.43\textwidth]{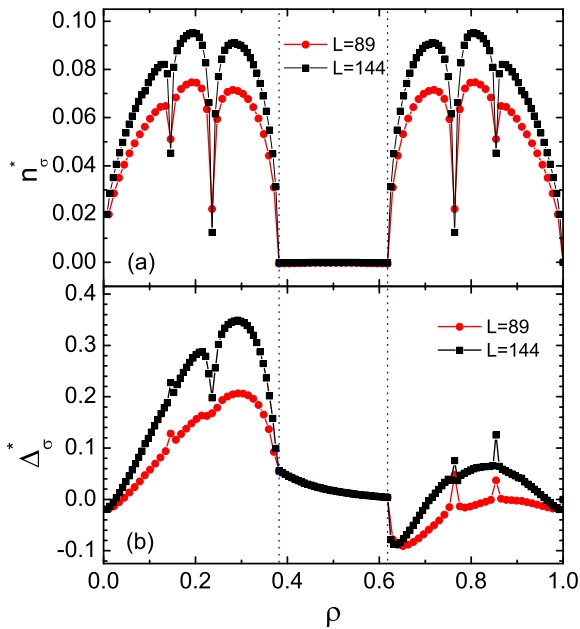}
\vspace{-0.3cm}
  \caption{(Color online) The primary sublattice peaks as a function of $\rho$ in (a) momentum distributions and (b) noise correlations,
  for two different system sizes, $\lambda=0.5$. Within the range from $\rho_1=1-\sigma$ to $\rho_2=\sigma$ (marked by blue
  dotted lines), the vanishing of peaks and the absence of scaling are noticed in $n^\ast_\sigma$ and $\Delta^\ast_\sigma$,
  respectively.}\label{fig:mispeak}
\end{figure}

As already discussed in the previous sections, there is an unexpected lack
of scaling with system size of the sublattice peak $\Delta^\ast_\sigma$ when $\rho=0.5$ in superfluid phase.
In Fig.\ \ref{fig:mispeak} we plot the sublattice peaks of the momentum distribution and the noise correlations
as a function of filling factor $\rho$ for $\lambda=0.5$, i.e., within the superfluid phase. We show results for
two system sizes to illustrate the scaling behavior with system size. Similar to $\Delta^\ast_\sigma$, we define
the primary sublattice peak in the momentum distribution as
$n^\ast_\sigma\equiv n_{\sigma\frac{2\pi}{a}}-\frac{1}{2}(n_{\sigma\frac{2\pi}{a}+\delta k}+n_{\sigma\frac{2\pi}{a}-\delta k})$,
with $\delta k=2\pi/La$. It is apparent in Fig.\ \ref{fig:mispeak}(a) that the sublattice peaks $n^\ast_\sigma$ are
absent in the window of filling factors between the fractional Mott states at incommensurate fillings
$\rho=1-\sigma$ and $\rho=\sigma$, respectively. Correspondingly, for $\Delta^\ast_\sigma$ [Fig.\ \ref{fig:mispeak}(b)],
a system size invariance is observed within the same filling window. The absence of scaling in the noise correlations
in that window of fillings then follows the absence of sublattice peaks in momentum distribution.

As mentioned in Sec. \ref{sec:local}, for $\lambda<\lambda_c$, the one-particle energy spectrum is split up by gaps
at incommensurate values of $k$ and the two largest gaps at $k=\pm\sigma\pi/a$ are associated with the fillings
$\rho=1-\sigma$ and $\rho=\sigma$, respectively. As a result, the fillings within the window $\rho=1-\sigma$ and $\rho=\sigma$
correspond to superfluid states with a partially-filled central band which is sandwiched by the two $\sigma$ gaps.
(There is actually a set of gaps in the central band, simply ignored as they are small compared to the $\sigma$ gaps.)
Interference effects in this window of fillings then seem to prevent those sublattice peaks from emerging in $n(k)$
and from scaling with system size in the noise correlations.

\begin{figure*}[!ht]
  \centering
  \includegraphics[width=0.70\textwidth]{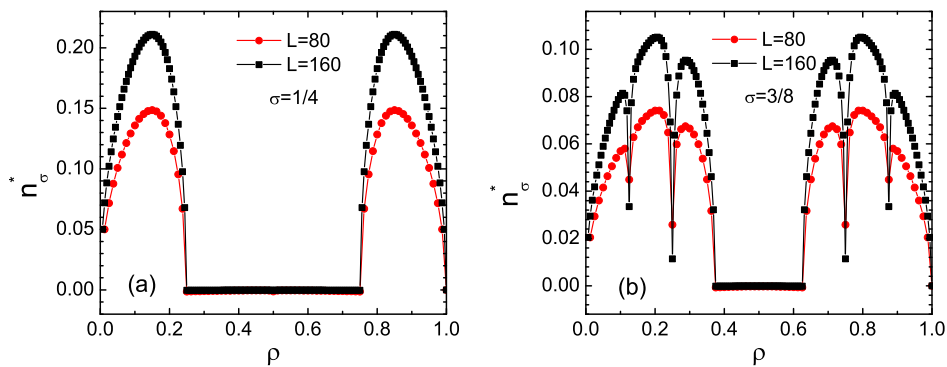}
\vspace{-0.2cm}
  \caption{(Color online) $n^\ast_\sigma$ as a function of filling factor in commensurate systems with period 4-multiples: (a) $\sigma=1/4$
  in a period-4 superlattice and (b) $\sigma=3/8$ in a period-8 superlattice. Two different sizes are plotted in both cases,
  for $\lambda=0.5$, which makes evident that the sublattice peaks are absent for fillings between
  $\rho=\sigma$ and $\rho=1-\sigma$.}\label{fig:mispeak2}
\end{figure*}

We should mention that a similar behavior can be seen in commensurate superlattices, but only in the case where the period
is a multiple of 4. Interestingly, for those periods, it was discussed in Ref.\ \cite{Rousseau06PRB} that the two central
bands in the spectrum cross at $k=0$. Results for the momentum and noise correlation peaks for commensurate superlattices
with period-$4$ and period-$8$ are presented in Fig.\ \ref{fig:mispeak2}. There, we plot $n^\ast_\sigma$ as a function of
$\rho$ for $\sigma=1/4$ and $\sigma=3/8$, respectively. The primary sublattice peaks in momentum distributions are
absent at fillings between 0.25 and 0.75 for $\sigma=1/4$ [Fig.\ \ref{fig:mispeak2}(a)] and between 0.375 and 0.625 for
$\sigma=3/8$ [Fig.\ \ref{fig:mispeak2}(b)]. Again, we attribute the missing peaks to a destructive
interference of the single particle wave functions, which may be related to the central band overlaps.\\

\section{Conclusions}\label{sec:conclusions}

Critical scaling exponents characterizing the behavior of correlation functions are the hallmark of critical phases and
phase transitions. In this paper, we have characterized the disorder-induced quantum phase transition between a superfluid
and a Bose-glass phase in two-color incommensurate superlattices by computing scaling exponents of the momentum
distribution and noise correlation peaks at $k=0$. At the quantum phase transition, those observables, which are accessible
in cold gases experiments, are found to scale with system size as $\sim N^{0.25}$ and $\sim N^{0.5}$, respectively. Intriguingly,
the values of the exponents are one-half of their corresponding values in the superfluid phase, where the system
is known to be described by the Luttinger liquid theory with Luttinger parameter $K=1$ \cite{review1D}. We also show that,
as expected, the scaling of those observables with system size vanishes in the localized phase.

We have studied the behavior of momentum and noise correlation peaks, as well as their first derivatives, as the
superlattice strength is varied. We have shown that both in HCB and noninteracting spinless fermion systems the
superfluid--Bose-glass transition is signaled by peaks in the first derivatives of the noise correlation. However,
only in the HCB case do those derivative peaks diverge with system size, providing a sharp experimental signature
of the phase transition.

\onecolumngrid

\twocolumngrid

\begin{acknowledgments}
K.H. and M.R. were supported by the Office of Naval Research. A.M.R. acknowledges support from an NSF-PFC grant.
I.I.S. is supported by ONR and NIST grants. We thank Edmond Orignac for useful discussions.
\end{acknowledgments}

\appendix
\section{Exact computational approach}\label{app:method}

In order to solve the HCB problem,
we map the Hamiltonian described in Eq.\ \eqref{eq:Ham} to the exactly solvable non-interacting fermion Hamiltonian
\begin{equation}\label{eq:HamFerm}
\hat{H}_\textrm{F} =-t \sum_{i} \left( \hat{f}^\dagger_{i}
\hat{f}^{}_{i+1} + \textrm{H.c.} \right)+ \sum_{i} V_i \,\hat{n}^f_{i},
\end{equation}
by means of a two-step process. The first step is to map HCBs to spin-1/2 systems through the
Holstein-Primakoff transformation (HPT) \cite{Holstein40}
\begin{equation}\label{eq:HPT}
\hat{\sigma}^+_{i} = \hat{b}^\dagger_{i}\; \sqrt{1-\hat{b}^\dagger_{i} \hat{b}^{}_{i}}, \quad
\hat{\sigma}^-_{i} = \sqrt{1-\hat{b}^\dagger_{i} \hat{b}^{}_{i}}\; \hat{b}^{}_{i},
\end{equation}
where $\hat{\sigma}^\pm_{i}$ are the spin raising and lowering operators for spin-1/2 systems.
The HPT leads to a replacement of $\hat{b}^\dagger_i (\hat{b}^{}_i)$ by $\hat{\sigma}^+_i (\hat{\sigma}^-_{i})$
if and only if the HCB creation and annihilation operators are normal ordered before the mapping.
This is a subtle point that needs to be considered for a correct calculation of the noise correlations if
one wants the HCB Hamiltonian to be the limit $U\rightarrow \infty$ of the Bose-Hubbard model.

Subsequently, the spin-1/2 Hamiltonian is mapped to the fermionic Hamiltonian by means of the Jordan-Wigner
transformation (JWT) \cite{Jordan28}
\begin{equation}\label{JWT}
\hat{\sigma}^{+}_i=\hat{f}^{\dag}_i
\prod^{i-1}_{\beta=1}e^{-i\pi \hat{f}^{\dag}_{\beta}\hat{f}^{}_{\beta}},\quad
\hat{\sigma}^-_i=\prod^{i-1}_{\beta=1} e^{i\pi \hat{f}^{\dag}_{\beta}\hat{f}^{}_{\beta}}\hat{f}_i,
\end{equation}
in which $\hat{f}^\dag_i$ ($\hat{f}^{}_i$) is the creation(annihilation) operators for spinless fermions.
In the noninteracting fermionic system the ground-state wave function can be expressed as a Slater determinant
\begin{equation}\label{eq:sd}
|\Psi_{F}\rangle=\prod^{N}_{\kappa=1} \sum^L_{\varrho=1} P_{\varrho \kappa}\hat{f}^{\dag}_{\varrho}\ |0 \rangle,
\end{equation}
with the matrix $({\bf P})_{L, N}$ given by the lowest $N$ single-particle eigenfunctions of the Hamiltonian
in Eq.\ \eqref{eq:HamFerm}. From the fermionic Slater determinants one can then calculate the two spin-1/2
Green's functions, $G_{ij}$ and $G_{ijkl}$:
\begin{eqnarray}
G_{ij}&=&\langle \hat{\sigma}^{-}_i\hat{\sigma}^+_j\rangle=\langle\Psi_{F}|\prod^{i-1}_{\beta=1}
e^{i\pi \hat{f}^{\dag}_{\beta}\hat{f}^{}_{\beta}} \hat{f}^{}_i \hat{f}^{\dag}_j
\prod^{j-1}_{\gamma=1} e^{-i\pi \hat{f}^{\dag}_{\gamma}\hat{f}^{}_{\gamma}} |\Psi_{F}\rangle\nonumber\\
&=&\det\left[ \left( {\bf P}^{i} \right)^{\dag}{\bf P}^{j}\right],
\end{eqnarray}
\begin{eqnarray}
G_{ijkl} &=& \langle \hat{\sigma}^-_i\hat{\sigma}^-_j\hat{\sigma}^+_k \hat{\sigma}^+_l\rangle =
\langle\Psi_{F}| \prod^{i-1}_{\alpha=1}e^{i\pi \hat{f}^{\dag}_{\alpha}\hat{f}^{}_{\alpha}} \hat{f}_i
\prod^{j-1}_{\beta=1}e^{i\pi \hat{f}^{\dag}_{\beta}\hat{f}^{}_{\beta}} \hat{f}_j \nonumber \\
&&\times \hat{f}^{\dag}_k \prod^{k-1}_{\gamma=1}e^{-i\pi \hat{f}^{\dag}_{\gamma}\hat{f}^{}_{\gamma}}
\hat{f}^{\dag}_l\prod^{l-1}_{\delta=1}e^{-i\pi \hat{f}^{\dag}_{\delta}\hat{f}^{}_{\delta}}| \Psi_{F}\rangle, \nonumber \\
&=& \det\left[ \left( {\bf P}^{ij} \right)^{\dag}{\bf P}^{kl}\right],
\end{eqnarray}
which are required to compute $n_k$ and $\Delta_{k k^\prime}$ in Eq.~\eqref{eq:defmom} \cite{He2011PRA}.
In the expressions above,
\begin{eqnarray}\label{eq:matcomp}
 P^{\alpha}_{\varrho \kappa}= \left\{ \begin{array}{rl}
 -P_{\varrho \kappa} & \text{for } \varrho<    \alpha,\,\kappa=1,\ldots,N \\
\,P_{\varrho \kappa} & \text{for } \varrho\geq \alpha,\,\kappa=1,\ldots,N \\
  \delta_{\alpha\varrho} & \text{for } \kappa=N + 1
\end{array}\right.
\end{eqnarray}
and
\begin{eqnarray}
 P^{\alpha\beta}_{\varrho \kappa}= \left\{ \begin{array}{rl}
 -P^{\beta}_{\varrho \kappa} & \text{for } \varrho<    \alpha,\,\kappa=1,\ldots,N+1 \\
\,P^{\beta}_{\varrho \kappa} & \text{for } \varrho\geq \alpha,\,\kappa=1,\ldots,N+1 \\
  \delta_{\alpha\varrho} & \text{for } \kappa=N + 2
\end{array}\right.
\end{eqnarray}
with $\alpha(\beta)=i,j,k,l$. The most time-consuming part of our calculations is determining
all the nonzero elements of $G_{ijkl}$,  each of which involves the multiplication of a $(N+2)\times L$
matrix and a $L\times (N+2)$ matrix, which scales as $(N+2)^2 L$, and then computing the determinant
of the resulting $(N+2)\times (N+2)$ matrix, which scales as $(N+2)^3$.

\end{document}